# El Bosón de Higgs (*)

Graciela B. Gelmini

*Department of Physics and Astronomy, University of California, Los Angeles (UCLA), 475 Portola Plaza, Los Angeles, CA 90095, USA. E-mail: gelmini@physics.ucla.edu*

**Resumen**: En el año 2012 fue descubierta la última partícula que completa el Modelo Estándar de las Partículas Elementales, la teoría mas sofisticada sobre la naturaleza en la historia de la humanidad. Si bien la presente formulación de la teoría proviene de los años 1960 y 70, ésta incorpora todo lo que miles de científicos descubrieron sobre partículas elementales y sus interacciones (excepto la gravitatoria) desde cerca de 1700 en adelante. Aunque brevemente, aquí trazamos la evolución de los principales conceptos incorporados en la teoría y explicamos la relevancia de la nueva partícula y del mecanismo por el que F. Englert y P. Higgs recibieron el Premio Nobel de física de 2013.

**Abstract**: The last particle that completes the "Standard Model of Elementary Particles", the most sophisticated theory of nature in human history, was discovered in 2012. Although the present formulation of the theory comes from the 1960's and 70's, it incorporates all discoveries that thousands of scientists made about elementary particles and their interactions (except for gravity) since the 1700's. Even if briefly, here we review the development of the major concepts included in the theory and explain the relevance of the new particle and the mechanism for which F. Englert and P. Higgs received the Nobel Prize in Physics 2013.

El 4 de julio de 2012 miles de físicos de partículas elementales se reunieron en auditorios distribuidos por todo el mundo, para presenciar un seminario transmitido a través de Internet en directo desde el auditorio principal de la Organización Europea para la Investigación Nuclear, CERN (Conseil Européen pour la Recherche Nucléaire) en Ginebra, Suiza, a las 9 hora local (fig. 1). En todos esos auditorios y por todo el mundo al unísono, los físicos mantuvieron la respiración en suspenso, gritaron exaltados y aplaudieron hasta que les dolieron las manos y algunos casi lloraron por la emoción. Entre los que sí lloraron estuvo el Dr. Peter Higgs, el físico inglés de 84 años que dio su nombre a una famosa partícula (figura 2). ¿Qué fue lo que hizo que estos físicos se comportaran como los hinchas en un partido de fútbol o el público de un concierto de rock? Lo que anunciaron en este seminario los portavoces de los dos experimentos principales del CERN, CMS y ATLAS, fue que ambos independientemente habían descubierto "una partícula compatible con" el "bosón de Higgs" [1], la última pieza del Modelo Estándar (ME) de las Partículas Elementales, formulado en los años 1960 y recién completado con este descubrimiento. Con la cautela propia de los científicos, sólo en marzo de 2013 [2], después de poder acumular datos suficientes para confirmar dos propiedades fundamentales de la partícula, declararon que ésta es "un bosón de Higgs", no aún "el bosón de Higgs", ya que llevará muchos años descartar o confirmar la posibilidad de encontrar otros "bosones de Higgs" no incluidos en el ME.

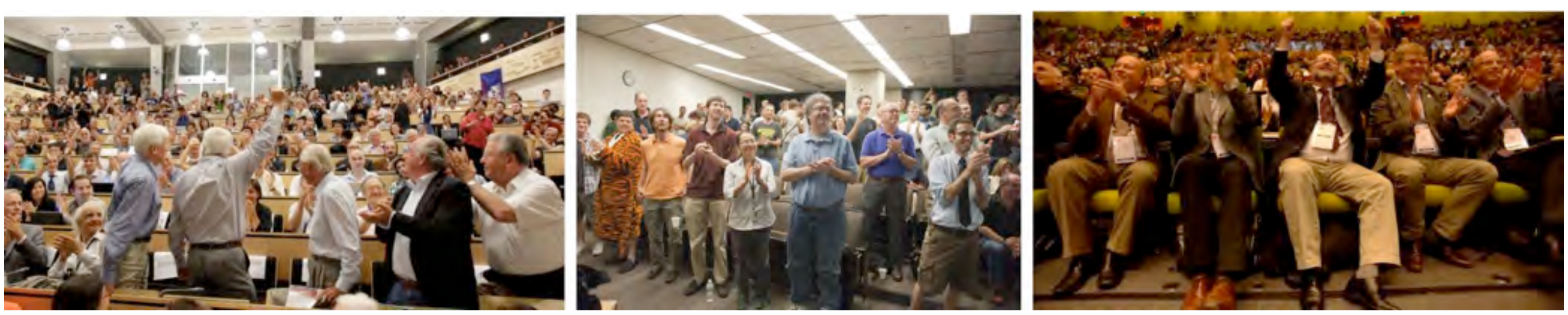

Figura 1: Reacción simultánea de físicos de partículas ante el anuncio del descubrimiento del bosón de Higgs. A la izquierda, en el auditorio principal del CERN, Ginebra, Suiza; en el centro, en un auditorio del Fermilab en Chicago EE.UU.; a la derecha, en un auditorio de Melbourne, Australia.

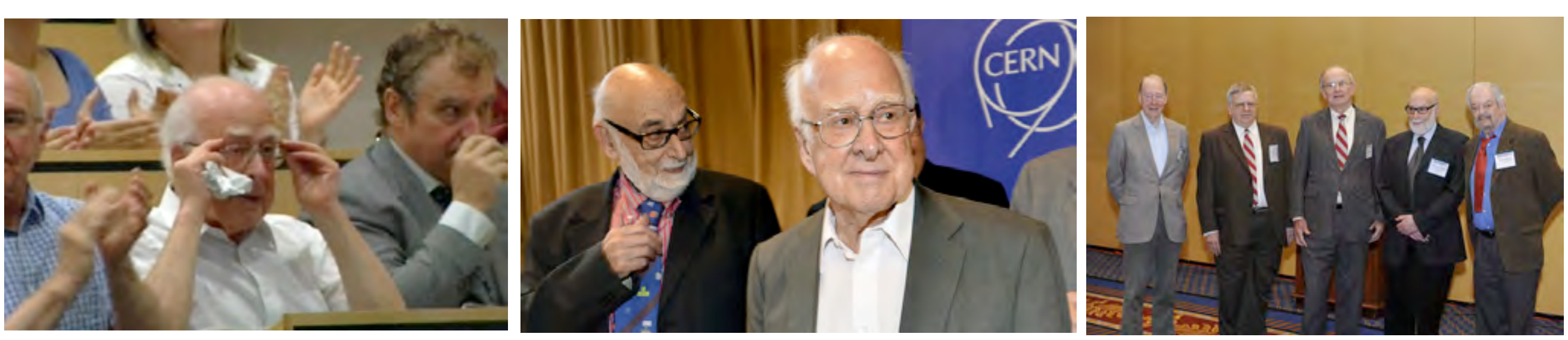

Figura 2: Izquierda: Peter Higgs muestra su emoción durante el anuncio del descubrimiento de la partícula que lleva su nombre. Centro: François Englert y Higgs después del anuncio de su Premio Nobel. Derecha: de izquierda a derecha, T. Kibble, G. Guralnik, C. R. Hagen, F. Englert, y R. Brout en 2010 (ver más adelante).

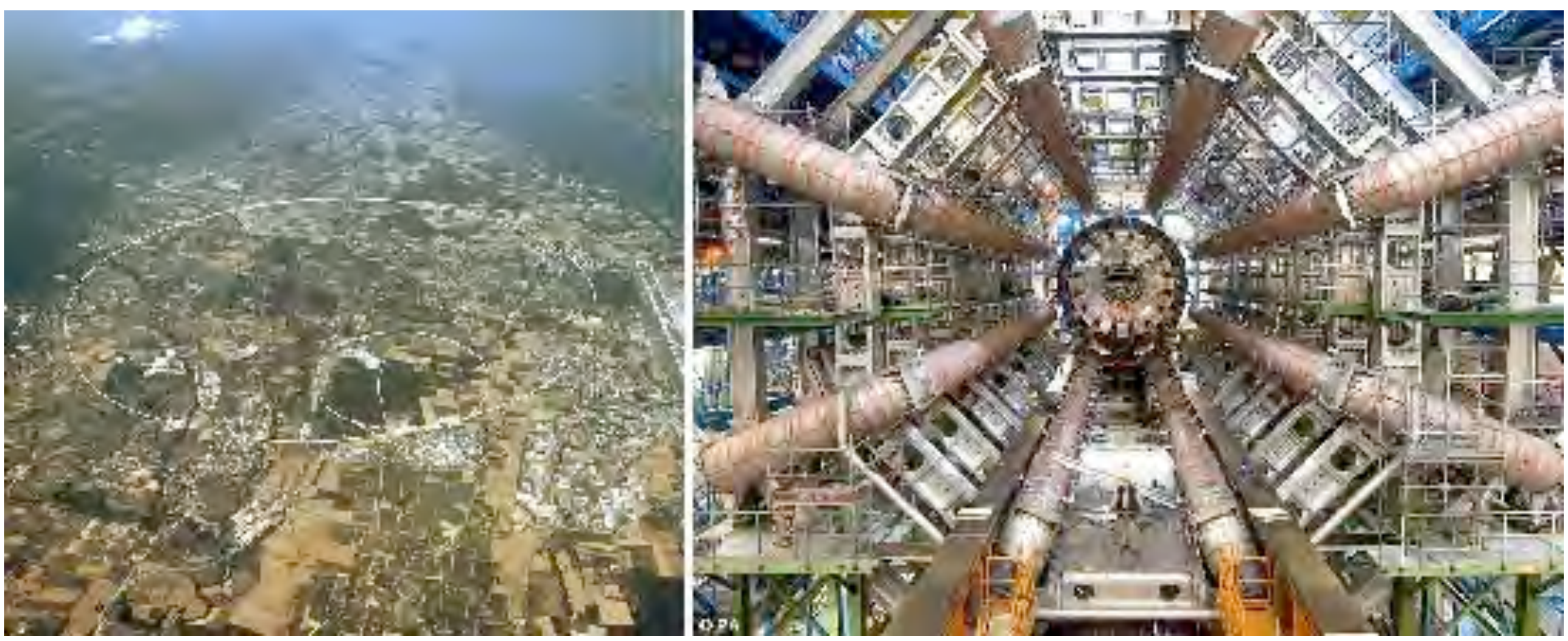

Figura 3: 3.a A la izquierda se ve el trazado del colisionador circular LHC (el círculo blanco más grande) superpuesto sobre una foto aérea de las afueras de la cuidad suiza de Ginebra, que muestra a la derecha parte de las pistas de aterrizaje del aeropuerto internacional de Ginebra. 3.b En la foto de la derecha se ve el detector del experimento ATLAS. Notar un hombre de pie que da una medida de la escala.

El LHC (Large Hadron Collider), Gran Colisionador de Hadrones, en el CERN [3], es el mayor colisionador de partículas del mundo, un acelerador circular de 27 kilómetros de largo (ver la figura 3.a), con varios detectores gigantes (como el de la figura 3.b) que registran las colisiones, y algunos consideran por su complejidad y magnitud el equivalente moderno de las catedrales de la edad media. El LHC fue construido, en gran medida, para encontrar el bosón de Higgs.

El 4 de Julio de 2012 el anuncio ocupó las portadas de los principales periódicos del mundo. Algunos (como La Nación, ver la figura 4) usaron el apodo "Partícula de Dios" [4], que muchos físicos consideramos desatinado por la confusión que genera y que proviene de un libro de divulgación escrito en 1993 por el Premio Nobel León Lederman. Su título, "The God Particle" (La Partícula de Dios) proviene en parte de acortar otro que Lederman quería por lo que estaba costando encontrarla, "The Goddamn Particle" (La Partícula Maldita), y que el editor no permitió.

Lo que sigue es un intento de explicar el ME para no-físicos. Esta teoría engloba la descripción de los constituyentes más fundamentales de la materia, los "leptones" y los "quarks", y

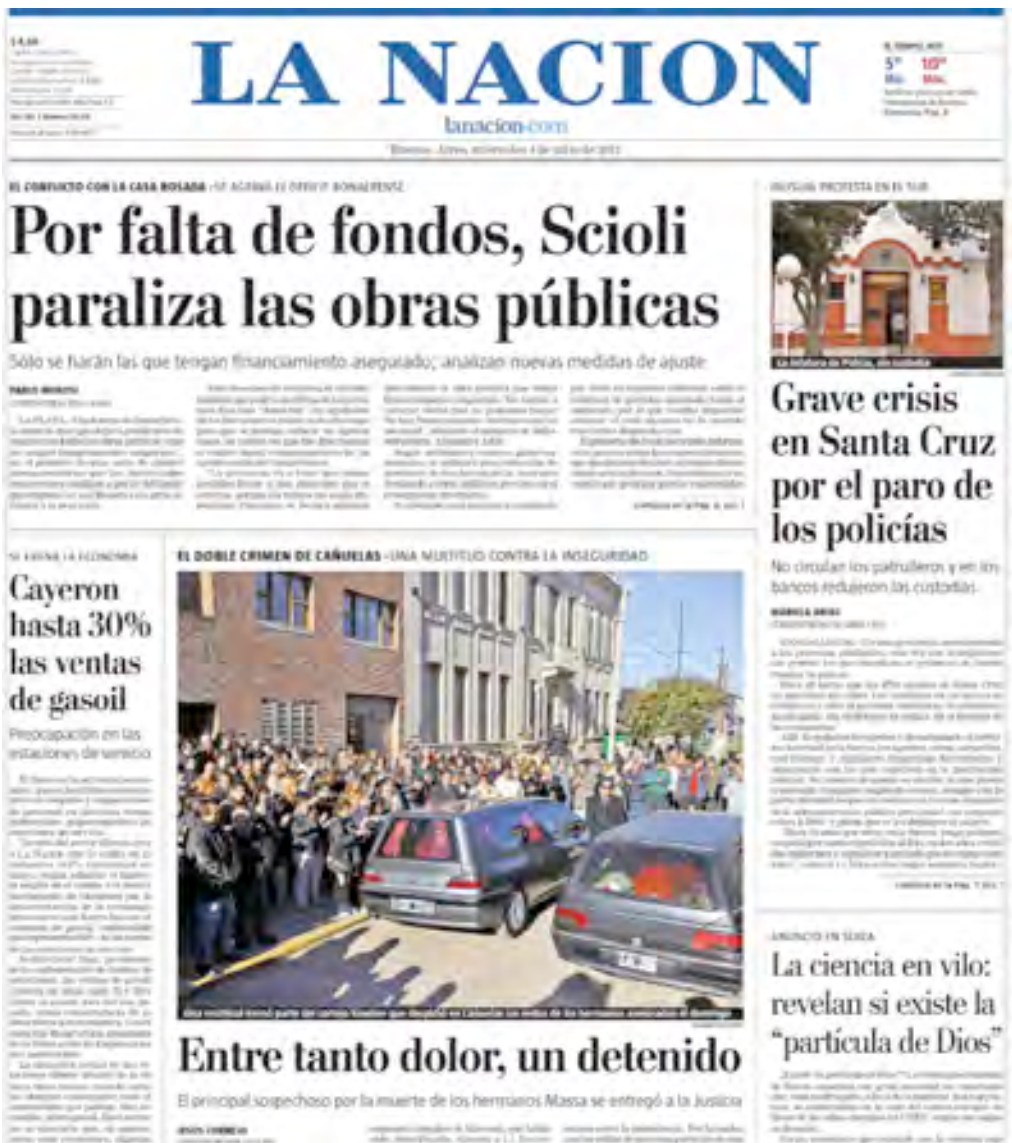

LA NACION

lanacion.com

Por falta de fondos, Scioli paraliza las obras públicas

Grave crisis en Santa Cruz por el paro de los policías

Cayeron hasta 30% las ventas de gasoil

Entre tanto dolor, un detenido

La ciencia en vilo: revelan si existe la "partícula de Dios"

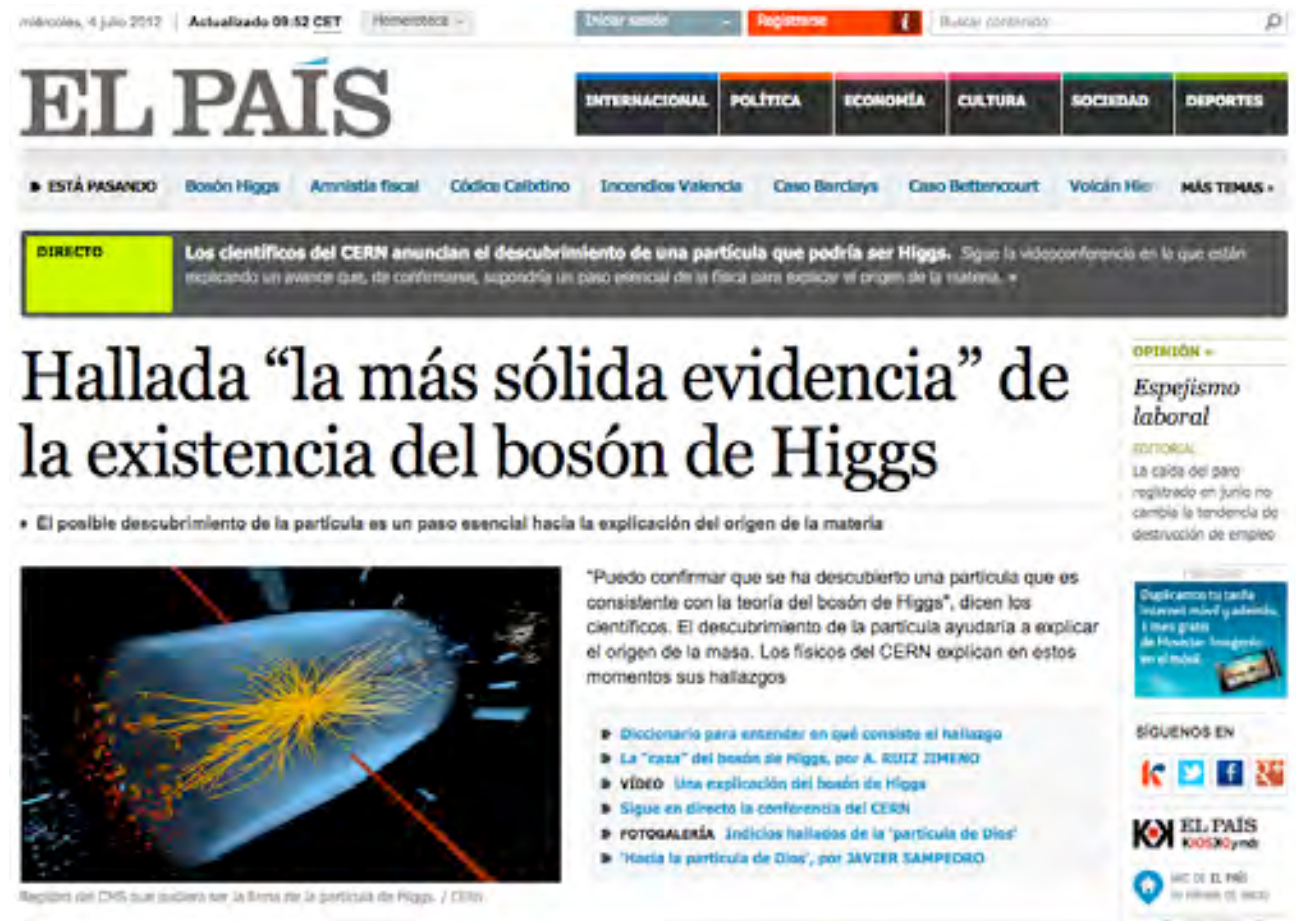

EL PAÍS

INTERNACIONAL POLÍTICA ECONOMÍA CULTURA SOCIEDAD DEPORTES

Los científicos del CERN anuncian el descubrimiento de una partícula que podría ser Higgs.

Hallada "la más sólida evidencia" de la existencia del bosón de Higgs

El posible descubrimiento de la partícula es un paso esencial hacia la explicación del origen de la materia

"Puedo confirmar que se ha descubierto una partícula que es consistente con la teoría del bosón de Higgs", dicen los científicos. El descubrimiento de la partícula ayudaría a explicar el origen de la masa. Los físicos del CERN explican en estos momentos sus hallazgos sus interacciones, a excepción de la gravedad, en términos de "campos cuánticos". Las interacciones fundamentales conocidas son cuatro: la gravitatoria, la electromagnética, la débil y la fuerte; el ME describe a la perfección las tres últimas. La interacción gravitatoria es despreciable con respecto a las otras a las distancias que exploran nuestros experimentos de laboratorio, lo que permite una descripción consistente de éstas sin incluirla. En el ME las interacciones electromagnéticas y débiles están unificadas en la interacción electrodébil y para ello la existencia del bosón de Higgs es esencial. Este es un ejemplo de una idea central en el desarrollo de la ciencia, la de "unificación", la idea de que fenómenos aparentemente independientes se deben a una única causa. Otra idea central en la ciencia es la de "simetría", que parece ser uno de los principios de la naturaleza, y está en la base del ME. Quizás, el mejor modo de entender los conceptos mencionados en este párrafo sea ver cómo y por qué se desarrollaron históricamente [5].

Figura 4: Portadas del diario argentino La Nación y del diario español El País del 4 de Julio de 2012

## Las primeras interacciones: gravitación y electromagnetismo

La gravitatoria es la más conocida de las interacciones y la que plantea mayores problemas teóricos ya que hasta ahora no se la ha podido describir consistentemente en términos de campos cuánticos (ver más adelante). Fue inicialmente descrita por Isaac Newton en 1687 en sus "*Principia*", en donde también introdujo el concepto de "fuerza" con su famosa ley "fuerza igual a masa por aceleración, $F = ma$". Newton probó que el movimiento de un proyectil en la Tierra y el de los planetas alrededor del Sol se deben a la misma fuerza. Con esto demostró que las leyes de la física que valen en la Tierra valen en todo el Universo. Este es el primer ejemplo de la poderosa idea de unificación. El siguiente es el del electromagnetismo.

Los fenómenos eléctricos y magnéticos datan por lo menos de siglo VI AC. Llaman poderosamente la atención porque son aparentemente "acciones a distancia" en las que un cuerpo ejerce una atracción o repulsión sobre otro sin tocarlo. Sólo en el siglo XVIII se reconoció (por obra de Charles de Fay y Benjamin Franklin) que hay dos tipos de carga, positiva y negativa, y que cargas del mismo tipo se repelen y de diferente tipo se atraen. Ahora sabemos que la carga eléctrica es una propiedad cuántica fundamental de las partículas elementales.

La electricidad y el magnetismo se consideraron fenómenos separados hasta que en 1820 H. C. Oersted descubriera que las corrientes eléctricas (o sea cargas en movimiento) producen fuerzas magnéticas y que en 1831 Michael Faraday descubriera que se puede generar una corriente eléctrica moviendo un imán dentro de un bucle de alambre conductor. Para explicar este efecto, Faraday visualizó "líneas de fuerza magnética" cerradas que emanan de uno de los polos de un imán y mueren en el otro polo (ver las figuras 5.a y 5.b), llenando todo el espacio en torno del imán y que generan corriente en un alambre que las corta. Esta descripción cambia profundamente el carácter de la interacción: el alambre interactúa localmente con las líneas y no a la distancia con el imán. Faraday también concibió "líneas de fuerza eléctrica" que se irradian de una carga eléctrica en todas direcciones (ver la figura 5.c y 5.d) y con las cuales cualquier otra carga interactúa localmente.

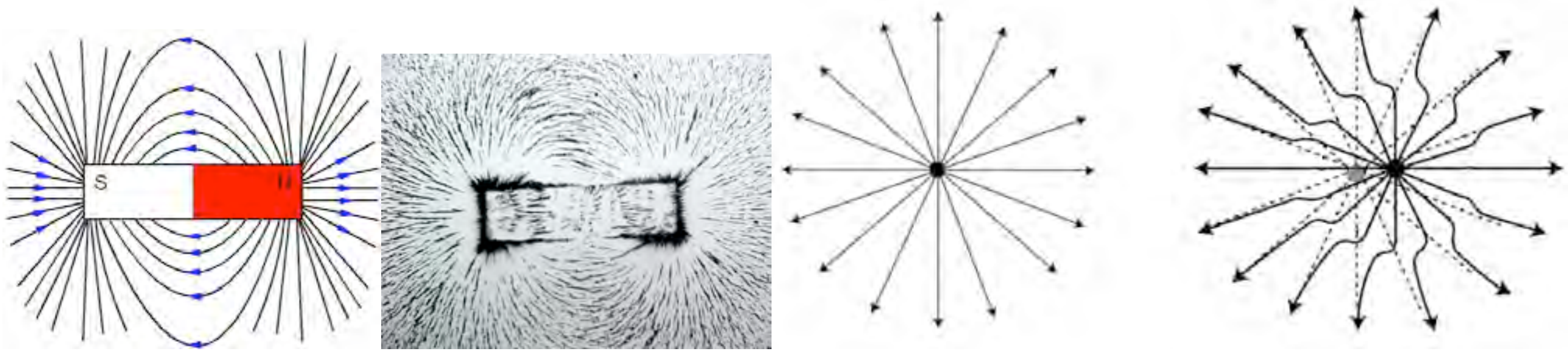

Figura 5: 5.a (izquierda) Líneas de campo magnético postuladas por Faraday conectando los dos polos de un imán que pueden ser vistas por la orientación que toman limaduras de hierro alrededor de un imán en la foto en 5.b (centro izquierda). Líneas de campo eléctrico saliendo de una carga positiva. 5.c (centro derecha) en reposo, y 5.d (derecha) cuando la carga es sacudida y una perturbación (onda electromagnética) empieza a propagarse por las líneas de campo.

### El concepto clásico de "campo" en el electromagnetismo: todas las interacciones son locales

Faraday inventó el concepto de "campo de fuerza", representado por las líneas que acabamos de ver. En física se dice que existe un "campo" de una cierta magnitud física cuando ésta adquiere en cada punto de un volumen un valor distinto, que puede cambiar con el tiempo.

La idea de un "campo de fuerza" es que la interacción a distancia de cuerpos se debe a que cada uno de ellos genera algo a su alrededor, un campo, que llena el espacio circundante y que interactúa con otro cuerpo en el lugar en que este otro está. Esto permite preservar el concepto de "localidad" de las interacciones, que corresponde a la intuición de que uno puede explicar el comportamiento de objetos cercanos sin referencia a otros lejanos, y la relación de causa y efecto en interacciones, ya que ninguna influencia puede ser instantánea (sino que viaja por el campo a una velocidad no mayor que la de la luz, lo que Faraday no sabía, pero que es un resultado central de la Relatividad Especial).

Faraday escribió cuatro ecuaciones que describen todos los fenómenos eléctricos y magnéticos, luego perfeccionadas por James Clerk Maxwell en 1865. Estas ecuaciones predicen la existencia de una onda en la que un campo eléctrico y uno magnético que varían en el tiempo se generan uno a otro sucesivamente y que puede viajar en vacío solo a una velocidad, la de la luz, usualmente denotada con la letra $c$. La luz es un tipo de estas "ondas electromagnéticas" que podemos detectar con nuestros ojos.

Estas ondas pueden visualizarse usando las líneas de fuerza de Faraday. Si sacudimos una carga, las líneas de fuerza que emanan de ésta transmiten la perturbación radialmente hacia afuera (ver la figura 5.d) como una onda, que al llegar a otra carga la sacude. La asociación que en Mecánica Cuántica se hace entre una onda y una partícula lleva entonces a interpretar la interacción entre dos partículas cargadas como el intercambio de otra partícula "de fuerza" (el "fotón" en este caso). La relación entre interacciones y sus partículas mediadoras es todavía más estrecha en el concepto de "campo cuántico" (que veremos mas adelante).

### El electromagnetismo y la Relatividad Especial

Ya Faraday entendía que una carga en reposo genera un campo puramente eléctrico y una en movimiento genera además uno magnético. Sin embargo, una misma carga puede parecerle estar en reposo a un observador pero en movimiento a otro que, por ejemplo, pase en un auto. O sea que un campo sea eléctrico o magnético depende de la velocidad del observador, por eso lo llamamos "electromagnético". Fue la necesidad de explicar el electromagnetismo lo que llevó a la Relatividad Especial, formulada por Albert Einstein en 1905.

El hecho de que las ecuaciones de Maxwell dan un solo valor para la velocidad de la luz en el vacío, $c$, constituye una de las bases de la Relatividad Especial: la velocidad de la luz es igual para cualquier observador (¡aún cuando éste se mueva con velocidades cercanas a $c$!) y para que esto sea así el concepto de espacio y de tiempo tiene que depender del observador. Esta fue una las ideas mas revolucionarias de principios del siglo pasado.

Entre las principales predicciones de la Relatividad Especial están: 1) la relación entre la masa $m$ y la energía $E$ en reposo de una partícula $E = mc^2$, 2) que una partícula con masa sólo puede viajar con una velocidad inferior a la de la luz y 3) que sólo una sin masa se mueve siempre a la velocidad $c$ (en el vacío). Para poder describir cualquier partícula, incluyendo una con $m = 0$, es necesario usar la forma completa y mucho menos conocida de la energía relativista: $E^2 = m^2c^4 + p^2c^2$, donde $p$ es el "momento" o "cantidad de movimiento". Para una partícula sin masa $E = pc$ sólo es energía de movimiento. Para un cuerpo con velocidad $v$ mucho menor que $c$, el momento fue definido por Newton como $p = mv$. El momento relativista, $p = mv/\sqrt{1 - (v^2/c^2)}$, se aparta de $mv$ a medida que $v$ se acerca a $c$. Las teorías relativistas coinciden con las teorías clásicas cuando todas las velocidades son mucho menores que $c$.

### Las primeras "partículas elementales": átomos, electrones, protones y neutrones

La idea de que toda la materia está compuesta de partículas elementales, o sea de partes no divisibles, data también al menos del siglo VI AC. La palabra "átomos" significa "indivisible" y viene del filósofo griego Demócrito. En 1803 el químico inglés John Dalton regresó a la misma idea para explicar los elementos, que en 1777 Antoine Lavoisier, "el padre de la química moderna", había distinguido de los compuestos.

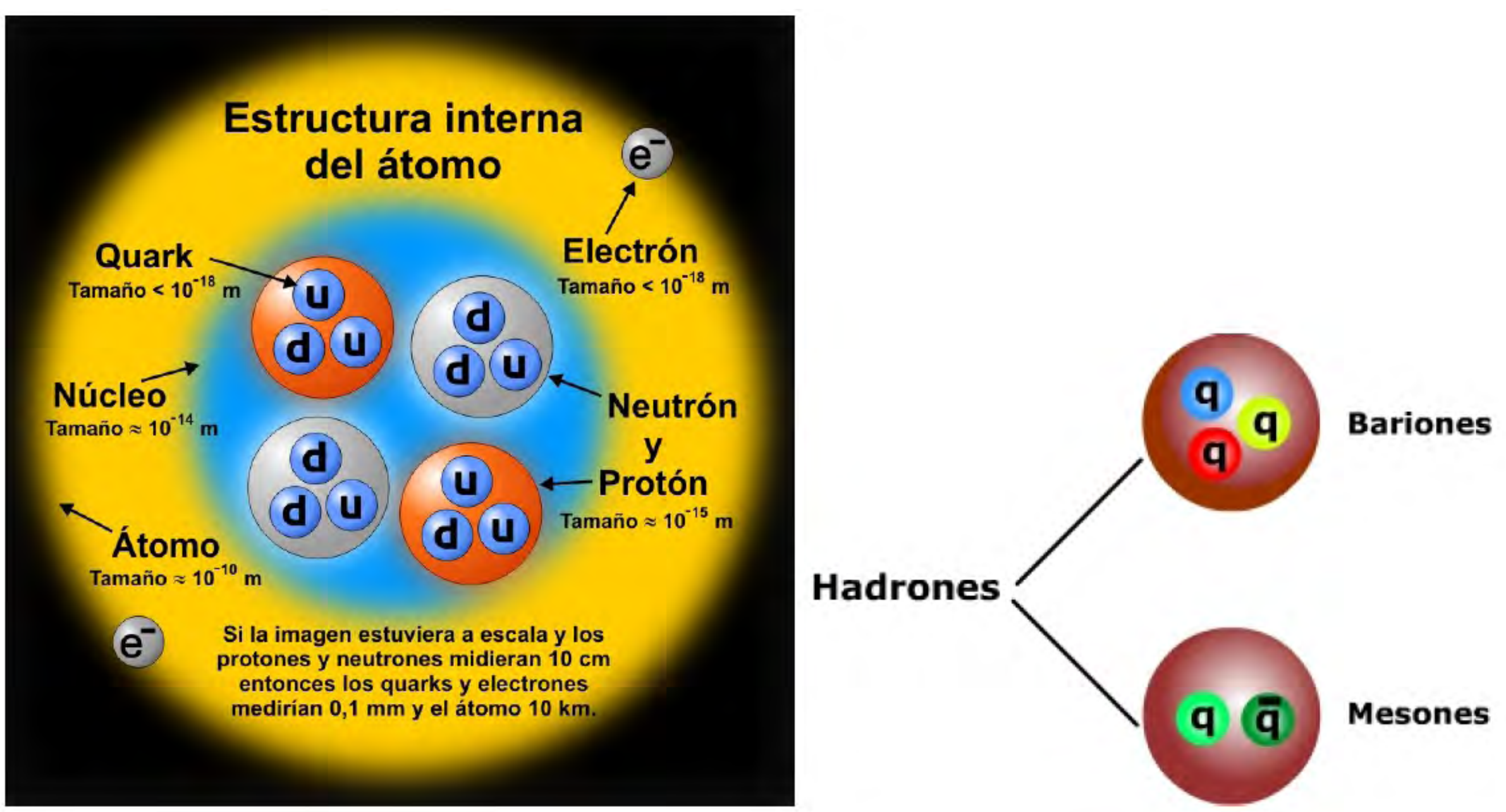

Figura 6: 6.a (izquierda) Esquema de la composición de un átomo mostrando su estructura en términos de las partículas elementales, quarks y electrones. 6.b (derecha) Esquema de la composición de los hadrones: los bariones, entre ellos los protones y neutrones, consisten de tres quarks, q, y los mesones de un quark, q, y un antiquark, $\overline{q}$ (una barra sobre el nombre de una partícula denota su antipartícula). Los colores representan la "carga de color" (que nada tiene que ver con el concepto usual de color) de los quarks y antiquarks, la fuente de la interacción fuerte. Todos lo hadrones tienen una carga total de color nula.

Sólo en 1911 los experimentos de Ernest Rutherford revelaron la verdadera estructura de un átomo: un "núcleo" compacto conteniendo toda la carga eléctrica positiva y prácticamente toda la masa del átomo y una nube de electrones que llevan toda la carga eléctrica negativa, de manera tal que el átomo es neutro (fig. 6). En un famoso experimento Rutherford lanzó núcleos de helio contra una lámina muy fina de oro, y vio que sólo unos pocos sufrían grandes deflexiones, indicando que habían chocado contra algo muy pesado, y los otros pasaban por el vasto vacío entre los núcleos de oro.

El electrón fue la primera partícula elemental descubierta (J. J. Thompson, 1897). Rutherford en 1920 dio el nombre de "protón" (del griego "protos", primero) al núcleo de hidrógeno (el más ligero), con carga eléctrica idéntica pero de signo opuesto a la del electrón, y postuló la existencia de protones y de partículas neutras, "neutrones", dentro de otros núcleos. James Chadwick descubrió el neutrón en 1932, con una masa similar a la del protón, unas 2000 veces la de un electrón. La masa de un átomo es casi igual a la suma de las masas de los protones y neutrones que contiene.

## Mecánica Cuántica y la dualidad onda-partícula

La Mecánica Cuántica, un nuevo tipo de mecánica que vale a escalas menores que $10^{-8}$ cm, fue propuesta para explicar el comportamiento de los átomos y sus constituyentes. Según la Mecánica Clásica los átomos no pueden ser estables porque los electrones deberían caerse al núcleo.

En 1900, para explicar el color que toman los materiales al calentarse, Max Planck postuló que la energía se emite y absorbe sólo en cantidades discretas, que él llamó "cuantos" de energía. En 1905, Albert Einstein (para explicar el efecto fotoeléctrico) propuso que la luz, aunque viaja como una onda, sólo puede ser absorbida o emitida como si consistiera de partículas que el llamó "cuantos de luz", rebautizados luego (G. Lewis, 1926) "fotones". La luz es a la vez una onda electromagnética y fotones. Este es el primer ejemplo de la "dualidad onda-partícula" en Mecánica Cuántica. En 1924 Louis De Broglie postuló que también las consideradas partículas, como el electrón, se comportan como ondas cuando se propagan (lo que fue confirmado experimentalmente en 1927).

La Mecánica Cuántica, iniciada por Niels Bohr en 1913, y formulada claramente por Werner Heisenberg en 1925 y Erwin Schrödinger en 1926 de dos maneras distintas equivalentes, nació de la necesidad de explicar por qué la energía es emitida y absorbida sólo en cantidades discretas y también la estabilidad de los átomos. La explicación de la Mecánica Cuántica es que cada estado de un sistema corresponde a una energía particular del mismo, y los valores intermedios están prohibidos, por lo que sólo puede pasar de un estado a otro emitiendo o absorbiendo una cantidad fija, o "cuanto", de energía. De modo que un electrón en un estado particular de un átomo es completamente estable, a menos que emita o absorba un cuanto de energía que le permita saltar a otro estado.

Una de las ideas completamente revolucionarias de la Mecánica Cuántica es que la posición de una partícula no puede determinarse con precisión. Lo único que puede determinarse es la probabilidad de encontrar a una partícula en un determinado lugar. Por ejemplo, la ecuación de Schrödinger aplicada a los electrones en un átomo tiene por solución una función, llamada "función de onda" o "estado", cuyo cuadrado es la probabilidad de encontrar los electrones en determinadas posiciones. El abandono del determinismo a las escalas subatómicas, tuvo consecuencias revolucionarias para el pensamiento científico y filosófico de principios del siglo XX.

## Bosones y fermiones

En mecánica clásica se puede determinar la posición y velocidad inicial de cada partícula con precisión y seguir su movimiento exactamente. De modo que la posición de cada partícula en cada instante es una etiqueta que permite diferenciar una partícula de cualquier otra. En Mecánica Cuántica, en cambio, la posición y velocidad de la partícula no pueden conocerse con precisión al mismo momento (este es el "principio de incerteza de Heisenberg"). Que la posición de cada partícula de un sistema sea incierta en Mecánica Cuántica implica que la posición no puede usarse para distinguir una partícula de otra del mismo tipo. Esta imposibilidad es una propiedad netamente cuántica que implica que la probabilidad no puede cambiar si se intercambian dos partículas del mismo tipo en el mismo sistema, ya que éstas son indistinguibles. Esto sólo es así si la función de onda o bien no cambia (en cuyo caso las partículas se llaman "bosones") o sólo cambia de signo (en cuyo caso las partículas se llaman "fermiones").

Los nombres "bosones" y "fermiones" derivan de los apellidos de Satyendra Nath Bose y Enrico Fermi, los físicos que exploraron las consecuencias de los respectivos tipos de partículas. En particular, dos fermiones, por ejemplo dos electrones, no pueden estar exactamente en el mismo lugar con las mismas propiedades. Es fácil ver esto en el caso frecuente en el que el estado de un sistema de partículas, como los electrones en un átomo, se puede describir como si cada una de ellas se moviera independientemente de las otras en un mismo campo de fuerza, como el campo eléctrico de un núcleo atómico, en estados de una partícula. Por ejemplo, la función de onda de dos electrones en un atómo, 1 y 2, puede escribirse usando la funciones de onda de cada uno, $f_1$ y $f_2$ como ($f_1 f_2$ - $f_2 f_1$) que cambia de signo al intercambiar 1 y 2, y es cero si $f_1$=$f_2$. Este es el "principio de exclusión de Pauli" (Wolfgang Pauli, 1925), que lleva a predecir todas las propiedades de la tabla periódica de los elementos. Por el contrario, no hay problema alguno en que todos los bosones estén en el mismo estado de una partícula.

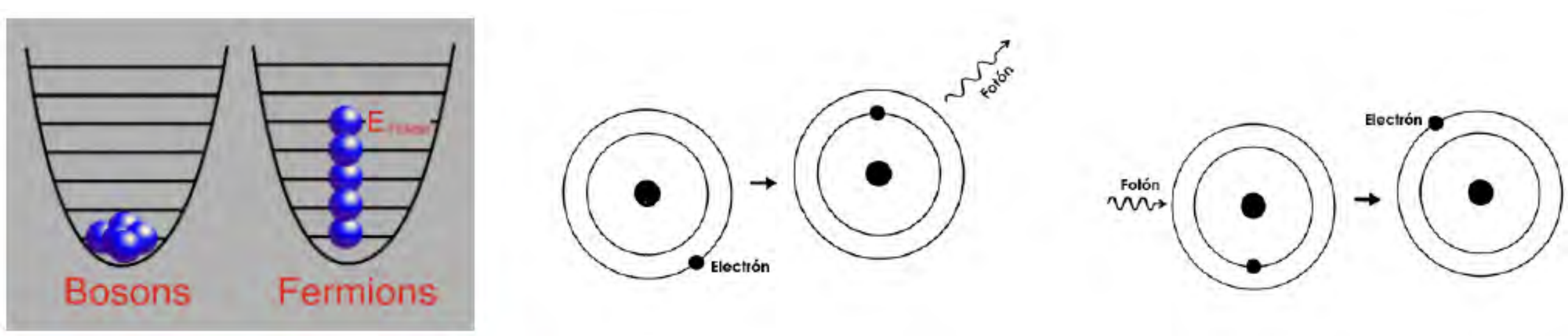

Figura 7: 7.a (izquierda) Bosones y fermiones en el estado de energía más baja de un sistema de partículas. Las líneas horizontales indican los diferentes niveles de energía de una partícula en presencia de un campo. Los bosones van todos al nivel de energía más bajo. En cambio, no más de un fermión puede estar en cada estado de una partícula. 7.b (centro) Emisión y 7.c (derecha) absorción de un fotón por un átomo al saltar uno de sus electrones de un nivel de energía a otro previamente vacío.

El estado más estable de todo sistema es el de menor energía. En éste, los fermiones ocupan uno a uno los estados de una partícula con energía más baja, mientras que los bosones van todos al estado de una partícula con menor energía, formando lo que se llama un "condensado de bosones" (ver la figura 7.a). Esta propiedad de los bosones es esencial en el rol que cumple el campo bosónico de Higgs en el ME.

## "Campos cuánticos", creación y destrucción de partículas y la Electrodinámica Cuántica

En el mundo macroscópico, las ondas electromagnéticas pueden generarse, por ejemplo, simplemente agitando una partícula cargada (ver la figura 5.d). En el mundo subatómico, la luz se emite de a un fotón a la vez. Un electrón en un átomo puede ir de un estado de energía a otro sea emitiendo o sea absorbiendo un fotón (ver las figuras 7.b y 7.c) y el fotón no existía antes de ser emitido y no existe después de ser absorbido.

En el mundo macroscópico, los cuerpos son permanentes. Si un tornillo se sale de un reloj, sabemos que el tornillo existía en el reloj antes de que saliera de él. En el mundo subatómico las partículas pueden ser creadas y destruidas. La equivalencia relativista entre energía y masa, tiene aquí su máxima expresión: una partícula puede convertirse en otras de distinta masa si recibe o da la energía que corresponde. Estos fenómenos sólo pueden entenderse combinando la Mecánica Cuántica con la Relatividad. El concepto de "campo cuántico" surge necesariamente de esta combinación, así como el de "antipartícula".

La ecuación de Schrödinger no es relativista y funciona bien para los electrones en un átomo porque estos se mueven a velocidades mucho menores que $c$. La Relatividad Especial fue incorpora da en la ecuacion de Oskar Klein y Walter Gordon (1926), que vale para bosones, y en la de Paul Dirac (1928) que vale para fermiones. Las soluciones de estas dos ecuaciones relativistas no admiten una interpretación probabilística como la solución de la ecuación de Schrödinger. Se deben interpretar, en cambio, como "operadores" (o sea, símbolos matemáticos que indican una operación sobre un objeto matemático) llamados "campos cuánticos". Estos crean o destruyen partículas en un estado sobre el que actúan, respetando las reglas de la Relatividad Especial. Esto es consecuencia de que la energía relativista $E$ puede tomar valores negativos, ya que $E^2 = m^2c^4 + p^2c^2$, aunque las partículas libres sólo tienen energía positiva. La existencia de antipartículas deriva de las soluciones con energía negativa.

Por ejemplo, la energía en reposo tiene dos soluciones, $E = mc^2$ y $E = - mc^2$, ya que $E^2 = m^2c^4$. Un campo cuántico caracterizado por un valor dado de la carga eléctrica y cargas de otro tipo, al actuar sobre el estado de un sistema adiciona estas cargas. Además, si tiene $E = - mc^2$ quita una energía $mc^2$ al aniquilar una partícula con energía positiva $mc^2$ (o sea que el signo menos significa que se resta una energía) y si tiene energía positiva $E = mc^2$, adiciona esa energía al crear una partícula con esa energía. En ambos casos el campo cuántico particular adiciona sus cargas, por lo que la partícula aniquilada debe tener cargas opuestas a las de la partícula creada por el mismo campo, pero la misma masa: la partícula que el campo aniquila es la "antipartícula" de la que crea. En 1932 Carl D. Anderson descubrió el anti-electrón, que llamo positrón, lo que confirmó experimentalmente la existencia de antipartículas.

Cada partícula tiene su antipartícula, con la misma masa y cargas opuestas. Partícula y antipartícula pueden coincidir cuando no llevan cargas que puedan distinguirlas. Por ejemplo, el fotón es su propia antipartícula, así como el bosón de Higgs.

La primera teoría de campos cuánticos, que sirvió de paradigma para todas las otras que vinieron después, fue la "Electrodinámica Cuántica" (QED, Quantum Electro-Dynamics), la teoría del electrón, el positrón (soluciones de la ecuación de Dirac) y el fotón (cuyo campo resulta de las ecuaciones de Maxwell). Después de los éxitos iniciales en los años 1930, la teoría encontró el problema de obtener valores infinitos al predecir ciertas cantidades observables. En los años 1940, fue desarrollado (principalmente por Shinichiro Tomonaga, Julian Schwinger y Richard Feynman) el proceso de "renormalización", que permite esconder esos infinitos en la definición de unas pocas constantes que se determinan experimentalmente. Después de esto, la QED ha probado ser una de las teorías más precisas que se han creado, capaz de hacer predicciones corroboradas experimentalmente de ciertas magnitudes físicas con una precisión de una parte en $10^{11}$.

La generalización matemática de la QED a interacciones con más de una partícula mediadora fue propuesta por Chen-Ning Yang y Robert Mills en 1954. Aunque su importancia no se reconoció en el momento, este tipo de teoría es la base del ME. Junto con la QED, estas se conocen como "Teorías de Gauge" y sus partículas mediadoras son bosones, llamados "bosones de gauge". "Gauge" significa "calibre" o "escala", y proviene de la primera tentativa, infructuosa, de formular la QED hecha por Hermann Weyl en 1919 aún cuando la teoría correcta que finalmente formuló en 1929 nada tiene que ver con calibres.

Para visualizar y describir las interacciones entre partículas, Feynman introdujo dibujos esquemáticos conocidos como "diagramas de Feynman" (fig. 8) en los que las partículas están representadas por líneas que se unen en los puntos de interacción. Las líneas rectas continuas representan fermiones (como el electrón, los "leptones" en general, y los "quarks") y las onduladas o discontinuas representan bosones (como el fotón, el pión y otros "mesones"). Estos diagramas tienen una interpretación matemática rigurosa, pero también representan las interacciones de manera intuitiva.

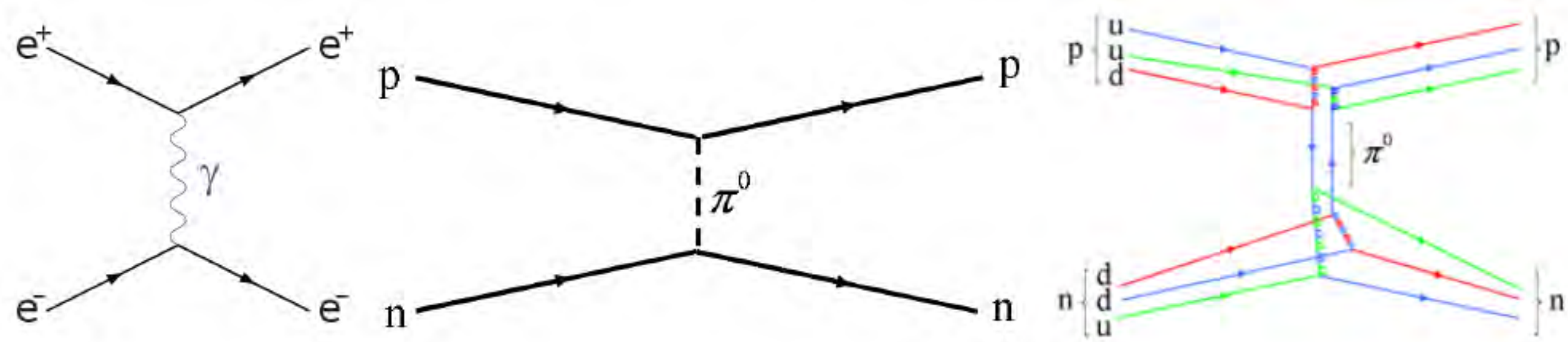

Figura 8: 8.a (izquierda) Diagrama de una interacción electromagnética entre un positrón (e+) y un electrón ($e^-$) por intercambio de un fotón (γ). 8.b (centro) Diagrama similar de la interacción entre un protón (*p*) y un neutrón (*n*) por intercambio de un pión neutro ($\pi^0$). 8.c (derecha) El mismo proceso de 8.b pero en términos de quarks (*u* y *d*) y gluones (representados con líneas de dos colores alternados); los colores de los quarks representan las cargas de color; los protones, neutrones y piones tienen carga de color total nula.

### Interacciones nucleares fuertes y débiles, neutrinos, piones y bosones $W^+$, $W^-$ y $Z^0$

Las interacciones electromagnéticas explican la estructura atómica pero para explicar la existencia y comportamiento de los núcleos atómicos son necesarias dos fuerzas fundamentales adicionales, las fuerzas nucleares "fuerte" y "débil".

Es evidente que debe haber una atracción que sienten ambos, neutrones y protones, y que los mantiene unidos formando un núcleo atómico. Ya que esta interacción debe ser más fuerte que la repulsión debida a la carga eléctrica de los protones, le fue dado el nombre de "interacción fuerte".

La "interacción débil" es responsable, por ejemplo, de la "radioactividad beta negativa", descubierta por Henri Becquerel en 1896. Este es un fenómeno (ver la figura 9.a) en el que un núcleo se transforma en otro de carga mayor, emitiendo un electrón. En términos de los constituyentes del núcleo, un neutrón se desintegra en un protón y un electrón. Sin embargo, algo no funciona bien con esta descripción: la energía del electrón emitido no toma un solo valor (dado por la diferencia de masa entre el núcleo original y el final) sino un continuo de valores. Esto llevó a Wolfgang Pauli en 1930 a hacer algo extremadamente audaz en su momento: postular la existencia de una nueva partícula emitida además del electrón y que por ser neutra no podía detectarse. En 1934 Enrico Fermi llamó "neutrino", o sea pequeño neutrón en italiano, a la partícula de Pauli, ya que su masa debía ser muchísimo menor que la de un electrón para describir bien los datos (como hizo Fermi en 1934, aunque el neutrino continuó eludiendo una detección directa hasta 1956).

Siguiendo el ejemplo del fotón, Hideki Yukawa propuso en 1935 la existencia una partícula, mediadora de la atracción entre protones y neutrones en un núcleo que él denominó "pi", *π* (ver la figura 8.b). Del tamaño de un núcleo, que indica el alcance de la fuerza fuerte, Yukawa infirió que su partícula debía ser más pesada que un electrón y menos que un protón y por esto la llamo "mesotrón" o "mesón" (del griego "mesos", medio) y ahora es llamado "mesón pi" o "pión".

El argumento usado por Yukawa es que el alcance *r* de una fuerza es inversamente proporcional a la masa *m* de la partícula mediadora de la fuerza. La explicación más simple de esta propiedad, dada por C. G. Wick en 1938, usa una de las relaciones de incerteza de la Mecánica Cuántica, la incerteza en la energía: el valor más probable de la diferencia $\Delta E$ entre dos mediciones de la energía de un sistema separadas por un intervalo de tiempo $\Delta t$ es igual a $h/(2\pi\, \Delta t)$, o sea $\Delta E \approx h/(2\pi\, \Delta t)$ (donde *h* es la "constante de Planck" $h/2\pi = 1{,}05\times10^{-34}$ Joules segundo). Esta relación tiene una implicación física importantísima: en un sistema cuántico la ley de la conservación de la energía puede no verificarse en un intervalo $\Delta t$ con un error menor que el $\Delta E$ dado por esta relación. En nuestro caso, durante el tiempo en el que una partícula mediadora de fuerza es intercambiada entre otras dos, hay tres partículas donde antes y después hay sólo dos sin que la energía total del sistema de dos o tres partículas haya cambiando. Esto es sólo posible si la incerteza en esta energía permite crear la partícula de masa $m$, o sea $\Delta E \geq mc^2$ durante un tiempo breve $\Delta t$ y en este tiempo la partícula puede moverse como máximo una distancia $r = c\ \Delta t$ lo que da el alcance de la interacción. De modo que $mc^2 \approx (h/2\pi)\ (c/r)$. Las partículas efímeras, que pueden existir sólo debido el principio de incerteza de la energía, se llaman “virtuales”.

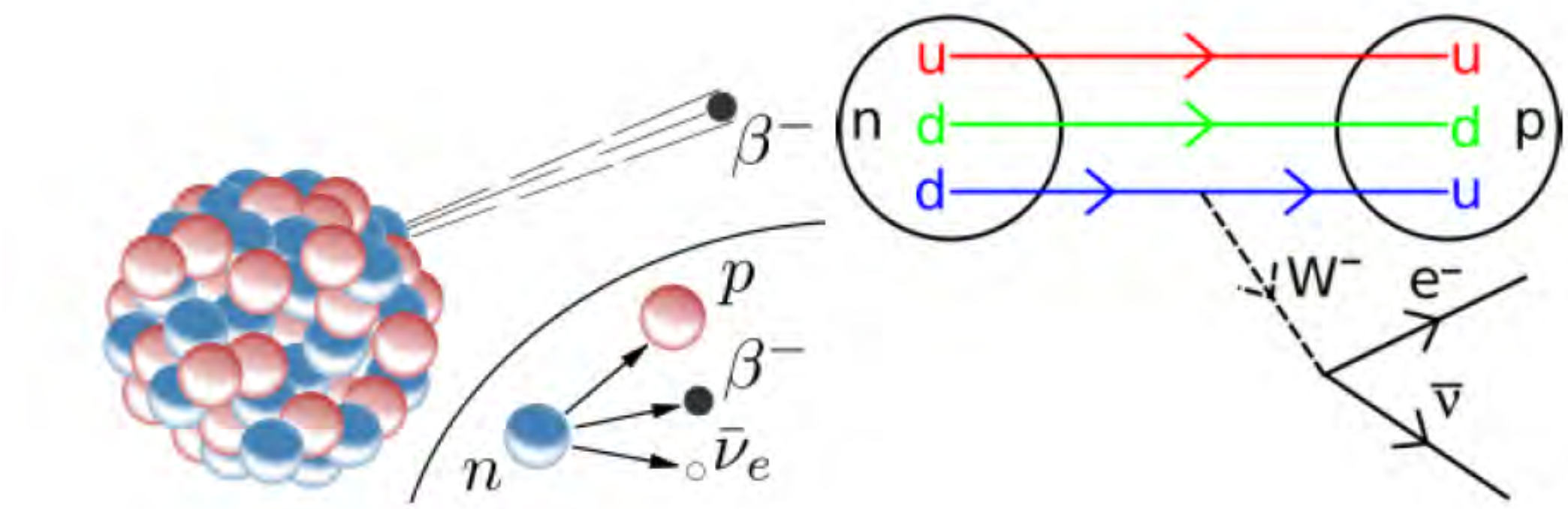

Figura 9: 9.a (izquierda) Desintegración beta negativa ($\beta^-$) de un núcleo y el diagrama correspondiente en términos de un neutrón y un protón. La partícula originalmente identificada como $\beta^-$ es un electrón. 9.b (derecha) Diagrama de Feynman de la misma desintegración vista al nivel de quarks. El tiempo corre hacia la derecha. Uno de los quarks $d$ del neutrón inicial emite un quark $u$ (lo que transforma el neutrón en un protón) y una partícula de fuerza débil $W$ que se desintegra en un electrón, $e^-$, y un antineutrino $\bar{\nu}$.

El fotón tiene masa cero lo que corresponde al alcance infinito de la interacción electromagnética. El pión, descubierto en 1947 por Cecil Powel, tiene una masa 273 veces la del electrón, lo que corresponde a un alcance de $2 \times 10^{-13}$cm, que es una fracción del tamaño de un núcleo (ver el esquema en la figura 6.a). A pesar del éxito de la descripción de Yukawa, aún quedaba mucho por hacer para entender las interacciones fuertes. En los años 1960 y 70 fue demostrado que los protones, neutrones y piones están constituidos de “quarks”, partículas que interactúan intercambiando “gluones” (ver la figura 8.c), como veremos más adelante.

La descripción inicial de Fermi en 1934 de la desintegración beta de un núcleo consistía en una interacción de contacto (o sea de alcance cero) de cuatro campos (ver la figura 9.a) que producían la destrucción de un neutrón y la creación en el mismo punto de un protón, un electrón y un antineutrino. Pero, una teoría con cuatro campos fermiónicos aplicados en el mismo punto produce resultados infinitos incurables (no es renormalizable). Eso llevó a Julian Schwinger en 1957 y luego a Sheldon Glashow, a proponer la existencia de partículas mediadoras de las interacciones débiles, llamadas $W^+$, $W^-$ y $Z^0$, que debían ser muy pesadas, del orden de 100 veces la masa de un protón, lo que corresponde a un alcance muy corto de $10^{-16}$ cm. Además, la descripción más fundamental de la desintegración beta debe hacerse en términos de quarks (como en la figura 9.b).

**Los leptones**

En 1936 Carl Anderson descubrió el “mu” ($\mu$) o muón, que es como el electrón, sólo que con masa 210 veces mas grande. La existencia de esta copia del electrón fue una total revelación, la primera indicación de que los leptones y quarks, vienen en tres copias que constituyen tres “familias” o “generaciones” (columnas I, II y III de la tabla en la figura 10), idénticas excepto por la masa que aumenta en cada una. Esta repetición, aún hoy no tiene explicación.

La primera partícula de la tercera generación, descubierta en 1975 (por Martin Perl), también fue una copia del electrón, el tau ($\tau$), cuya masa es unas 4000 veces las del electrón.

Las partículas que no tienen interacciones fuertes fueron llamadas (por Leon Rosenfeld en 1948) “leptones”, inicialmente para indicar que eran ligeras, del griego “leptos”, delgado (ver la figura 11.a). Ahora este grupo incluye el electrón, $e$, el $\mu$, el $\tau$, y sus correspondientes neutrinos: el del electrón $\nu_e$, el del muón, $\nu_\mu$ y el del tau, $\nu_\tau$ (descubiertos en 1956, 1962 y 2000 respectivamente).

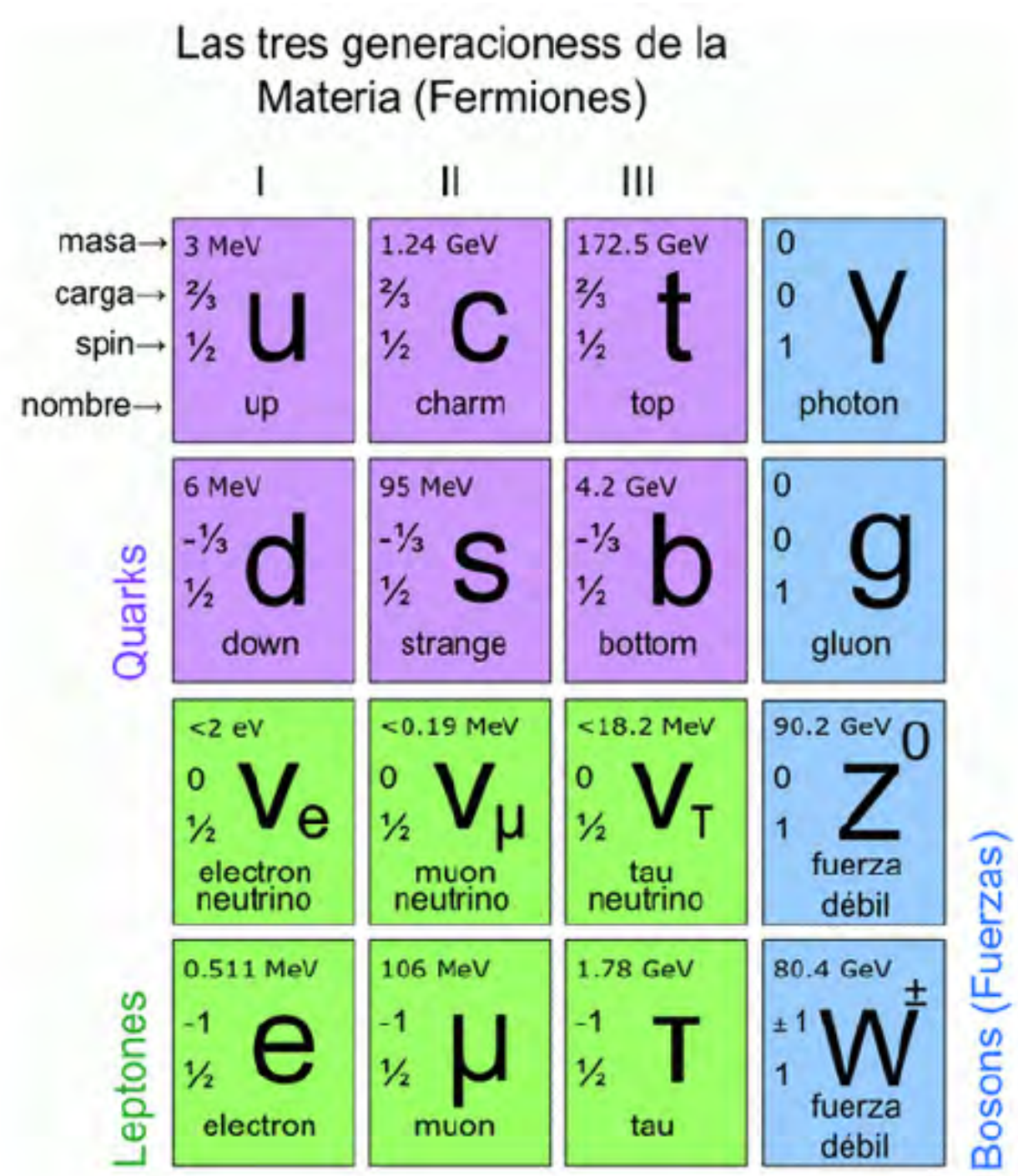

Figura 10: Tabla de las partículas elementales, con las familias o generaciones de leptones y quarks (que son fermiones) dadas en las primeras tres columnas, y las partículas mediadoras de fuerzas (que son bosones) dadas en la cuarta columna. La masa de un electrón es 0,5 MeV (aquí se toma $c$=1), 1 MeV = 1000000 eV y 1 GeV = 1000 MeV. La carga eléctrica y el nombre de cada partícula están incluidos. El “espin” o “spin” es una propiedad cuántica de las partículas que no hemos mencionado. En esta tabla falta sólo el bosón de Higgs.

## De los hadrones a los quarks

Los neutrones y protones fueron llamados “bariones” (del griego “varys”, pesado) porque eran las partículas más pesadas conocidas hasta el año 1947 (ver la figura 11.a). En 1962, Lev Okun llamó colectivamente “hadrones” (del griego “hadros”, denso) a las partículas que tienen interacciones fuertes, que son los mesones y los bariones.

El desarrollo de nuevos aceleradores y detectores de partículas en la década de los 1950 llevó al descubrimiento de un enorme número de hadrones (ver la figura 11.b), de masa cada vez más grande, pero que se desintegran muy rápidamente, en $10^{-8}$ segundos o menos. Ante la explosión del número de partículas se atribuye a Wolfgang Pauli la exclamación “Si hubiera previsto esto me hubiera hecho botánico”.

En 1961, Murray Gell-Mann e independientemente Yuval Ne'emann encontraron un esquema de clasificación de los hadrones en conjuntos de masa casi igual y conteniendo 8 o 10 partículas. Este fue el preludio de la proposición revolucionaria de Gell-Mann y George Zweig de tres años más tarde. En 1964 ellos explicaron la clasificación de los hadrones suponiendo que todos ellos consisten de partículas que Gell-Mann llamo “quarks” (ver la figura 6.b). En ese momento tres quarks (u, d y s) eran suficientes para formar todos los hadrones conocidos y el muy erudito Gell-Mann se inspiro para elegir el nombre “quark” (que en inglés es una imitación del grito de una gaviota) en una asociación libre con la frase “*Three quarks for Muster Mark*” (Tres quarks para el Señor Mark) en un poema de James Joyce en el que usa palabras que recuerdan sonidos de pájaros.

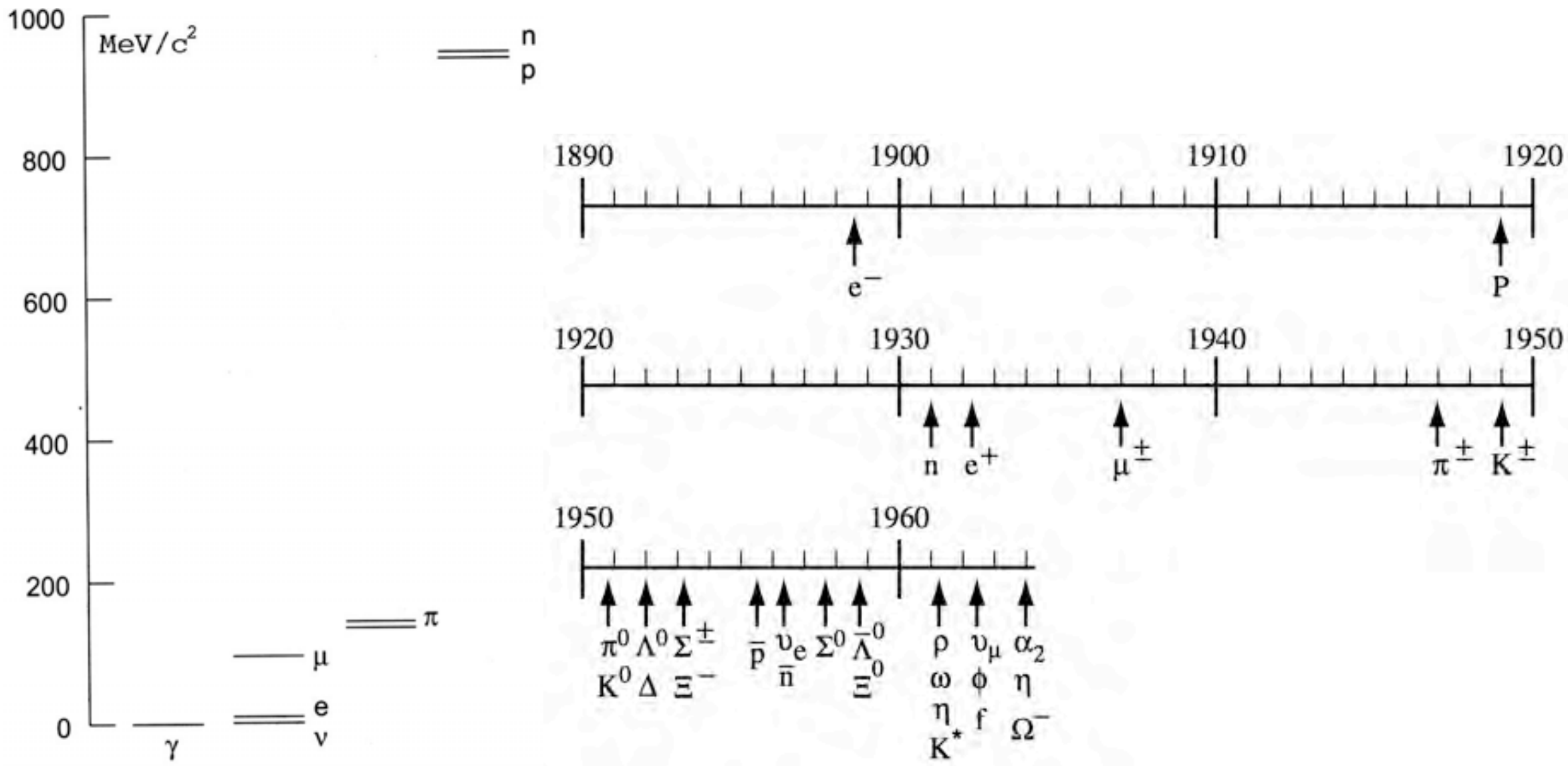

Figura 11: 11.a (izquierda) Partículas conocidas hasta 1947 y sus masas, dadas en la escala vertical en unidades de MeV/$c^2$ en las que la masa del electrón es 0.5. Las columnas de izquierda a derecha muestran respectivamente el fotón, los leptones (electrón, muon y neutrino), los mesones (sólo piones) y los bariones (el neutrón y el protón). Los mesones y bariones se llaman colectivamente hadrones (sienten las interacciones fuertes). 11.b (derecha) Línea de tiempo del descubrimiento de partículas hasta 1964 (año de la propuesta de los quarks como constituyentes de todos los hadrones). Una barra encima del nombre de una partícula denota su antipartícula (e.g. $\overline{p}$ es el antiprotón) o también la indicación de la carga eléctrica denota la partícula y la correspondiente antipartícula (e.g. $e^-$ es el electrón y $e^+$ es el positrón). Excepto por dos neutrinos, $\nu_e$ y $\nu_\mu$, todas las partículas que se muestran después de 1946 son hadrones.

Inicialmente vistos sólo como una manera de clasificar hadrones, como si fueran cubos para armar, los quarks fueron encontrados como partículas en una versión moderna del experimento de Rutherford que probó la existencia de núcleos dentro de los átomos. Protones y neutrones fueron bombardeados con electrones en experimentos realizados entre 1967 y 1971 que confirmaron que los quarks son los constituyentes de todos los hadrones y son, por lo tanto, las verdaderas partículas elementales que sienten las interacciones fuertes. Sin embrago, los quarks no se encuentran jamás afuera de un hadrón. Los quarks tienen cargas eléctricas fraccionarias, el *u* tiene carga 2/3 y el *d*, -1/3, y jamás se ha observado una carga fraccionaria libre.

Ahora se sabe que los quarks son seis, distinguidos por sus así llamados “sabores” *u, d, s, c, b* y *t*. El “up” (en inglés “arriba”) quark *u* y el “down” (“abajo”) *d* pertenecen a la primera genera_ ción. El “charm” (“encanto”) quark *c* y el “strange” (“extraño”) *s* y luego el “top” (“superior”) quark *t* y el “bottom” (“inferior”) *b,* son las repeticiones de los dos primeros que están respectivamente en las generaciones II y III (ver la tabla en la fig. 10).

En términos de quarks la desintegración beta negativa de un núcleo consiste de un quark *d* en un neutrón que emite un quark *u* (lo que transforma el neutrón en un protón) y un bosón débil de carga negativa $W^-$, que luego se desintegra en un electrón, $e^-$, y un antineutrino $\overline{\nu}$ (ver la figura 9.b). Los quarks de las generaciones II y III sufren desintegraciones débiles muy rápidas (que los llevan a los quarks estables de la primera generación), de modo que los hadrones que los contienen tienen una vida muy breve. El quark mas pesado, el *t*, encontrado experimentalmente recién en 1995 en el Fermilab, tiene una masa 175 veces la de un protón y su vida es tan breve, $5 \times 10^{-25}$ segundos, que es el único que no tiene tiempo de formar un hadrón antes de desintegrarse.

A pesar del enorme éxito del modelo de quarks, en 1964 todavía había que explicar por qué sólo algunas combinaciones de varios quarks (sólo sistemas de quark-antiquark y de tres quarks) aparecen en la naturaleza y no otras. En 1964 Oscar Greenberg y luego Moo Young Han y Yoichiro Nambu en 1965, propusieron la existencia de una carga adicional de los quarks (llamada mas tarde "color"), tal que sólo sistemas con valor cero de esta carga deberían existir en la naturaleza (para lo cual esta debía tener tres valores posibles para cada quark). No se tuvo idea de como explicar esta propiedad hasta los años 70.

## El Modelo Estándar de las Partículas Elementales

La palabra "modelo" en el nombre viene del periodo de los años 1970 en el que todavía la teoría no había sido suficientemente probada. Ahora el ME es una teoría comprobada experimentalmente en cientos de modos distintos.

El ME es una teoría relativista de campos cuánticos, basada en las ideas de unificación y de simetría que incorpora todos los elementos que hemos presentado hasta ahora. Es la teoría de todas las partículas elementales puntuales que conocemos (o sea que no muestran ninguna estructura interna, que son los leptones y los quarks en la tabla en la figura 10), sus respectivas antipartículas, y los bosones mediadores de fuerza conocidos (en la misma tabla).

Ya vimos varios ejemplos previos de la idea de unificación. En el ME las interacciones electromagnéticas y débiles están unificadas en la teoría electrodébil de Glashow, Weinberg y Salam y la teoría de la interacción fuerte es la Cromodinámica Cuántica. Ambas son teorías de Yang y Mills.

En física una simetría [6] de un objeto es una transformación que lo deja "invariante", o sea idéntico a sí mismo. Por ejemplo, si rotamos un cuadrado 90 grados alrededor de su centro, seguiremos teniendo un cuadrado idéntico al inicial. Esta es una transformación "discreta" en la que el parámetro, el ángulo, que la define sólo puede tomar ciertos valores. Un círculo es, en cambio, invariante bajo una rotación cualquiera alrededor de su centro. Esta es una transformación "continua", ya que el parámetro que la define, el ángulo de rotación, puede tomar cualquier valor.

La tremenda importancia de la simetría en física reside en que por cada simetría continua de una ley física hay una cantidad física conservada. Este teorema demostrado por la matemática Emmy Nöther en 1918 es un resultado central en física teórica. Por ejemplo, el hecho de que un experimento de laboratorio dé el mismo resultado hoy, que mañana, que dentro un siglo, o sea que las leyes de física que lo determinan sean invariantes respecto de traslaciones temporales, asegura la conservación de la energía.

Hay transformaciones, llamadas "internas", que no involucran traslaciones o rotaciones en el espacio o el tiempo. Una de estas simetrías (la multiplicación de campos por una fase) lleva a la conservación de la carga eléctrica (la suma total de las cargas participantes es la misma antes, durante, y después de todo proceso) y a la QED. La simetría bajo rotaciones de varios campos entre sí lleva a las teorías de Yang y Mills.

## La interacción electrodébil en el Modelo Estándar: la teoría de Glashow, Weinberg y Salam

La unificación del electromagnetismo y las interacciones débiles proviene de la simetría de las interacciones bajo rotaciones simultáneas de las partículas interactuantes entre sí y de las partículas mediadoras entre sí, relacionadas de manera matemáticamente precisa en las teorías de Yang y Mills. Por ejemplo, tomando como punto de partida la interacción de un electrón que va a un fotón y a un electrón (mostrada en la figura 12.a) y transformando simultáneamente el electrón inicial en un neutrino (del electrón) y el fotón en un $W+$, obtenemos otra interacción (mostrada en la figura 12.b) que se realiza en la naturaleza y forma también parte de la teoría.

Estas relaciones llevaron primero a Julian Schwinger en 1957 y luego a su ex-estudiante Sheldon Glashow en 1961 a proponer que las interacciones electromagnéticas, mediadas por el fotón, y las débiles, mediadas por los bosones cargados $W+$ y $W^-$, se pueden describir conjuntamente como parte de una sola teoría. Por motivos de consistencia del modelo, Glashow agregó otra partícula mediadora de interacciones débiles, el bosón neutro $Z^0$. Pero, la simetría (llamada "simetría de gauge") en la teorías de Yang y Mills impone que los bosones mediadores (los "bosones de gauge") no tengan masa. Además en el modelo de Glashow la misma simetría también impone que los leptones y los quarks tengan masa nula. La solución de este problema propuesta en 1964 es el "mecanismo de Higgs" que, como veremos en detalle, consiste en que la simetría de la teoría esté "espontáneamente rota" (aunque la simetría no está en verdad rota, sino que sólo no es aparente).

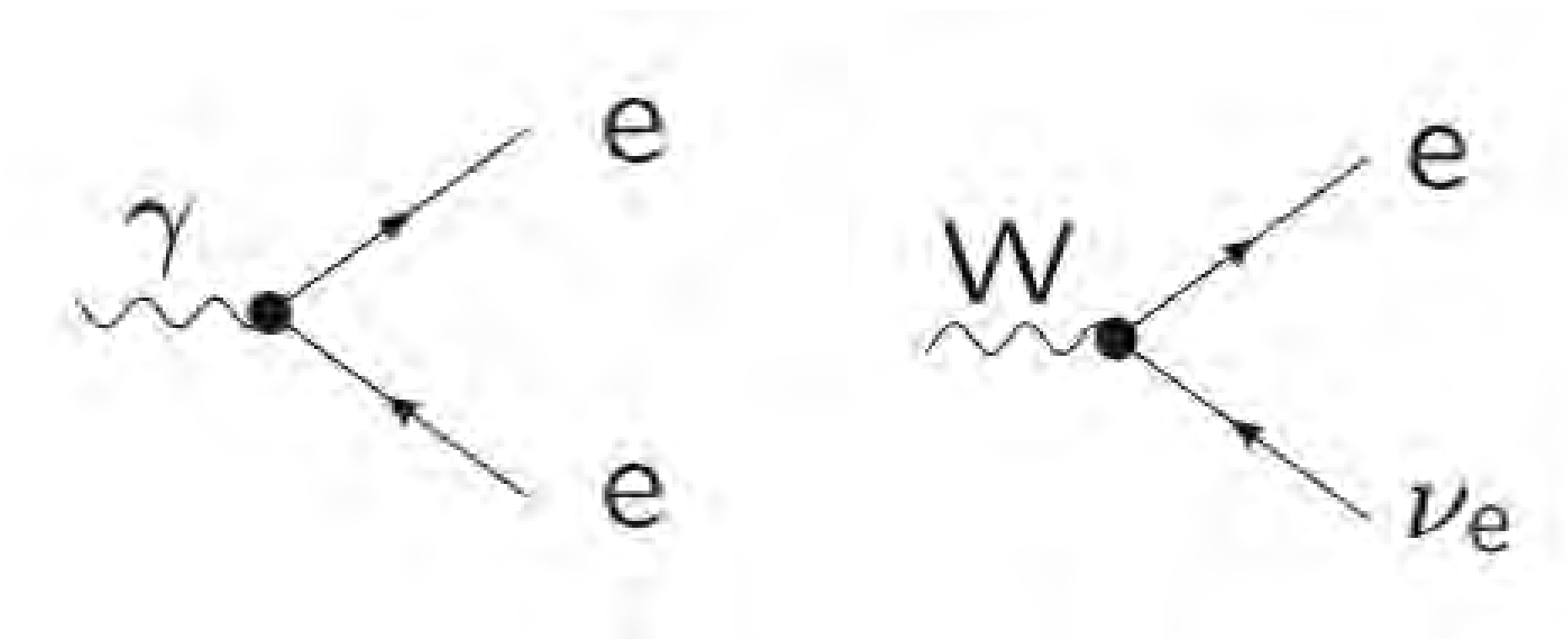

Figura 12: 12.a Diagrama de una interacción electromagnética en la que un electrón, *e*, emite otro electrón, *e*, y un fotón, *γ*. 12.b Diagrama similar de una interacción débil en la que un neutrino $\nu_e$ emite un electrón, *e*, y un bosón de carga positiva W+. Se puede pasar del primero al segundo cambiando el electrón inicial por un neutrino (del electrón) y simultáneamente el fotón por un W+.

Este mecanismo fue aplicado al modelo de Glashow por Steven Weinberg en 1967, e independientemente por Abdus Salam en 1968, quienes, postulando la existencia de una nueva partícula elemental, el bosón de Higgs, y su respectivo campo cuántico, lograron dar su masa a los mediadores de la interacción débil, W+, $W^-$ y $Z^0$ y a los leptones y quarks, dejando el fotón sin masa, como debe ser, lo que completó la teoría electrodébil.

En 1970 hubo un avance decisivo para el ME: Martinus Veltman y Gerard 't Hooft probaron que las de Yang y Mills son teorías de campos consistentes, o sea renormalizables, y en 1971 't Hooft extendió la prueba a aquéllas con masas generadas por el mecanismo de Higgs.

Uno podría suponer que el artículo de Weinberg de 1967, tuvo inmediatamente cientos de citas. Pero esto no fue así. Entre 1968 y 1970 el artículo tuvo sólo una cita por año (en 1968, en lecciones de Salam). El motivo de la casi total falta de interés inicial en la teoría electrodébil fue que esta no parecía ser válida para los quarks, o sea los hadrones. Este grave problema fue solucionado en 1970 cuando Sheldon Glashow, John Iliopoulos y Luciano Maiani reconocieron la importancia crítica de la existencia del quark *c* aún no descubierto al momento, y predijeron su masa. Sólo después de que la existencia de las interacciones del $Z^0$ fuera probada experimentalmente en el CERN en 1973, y el quark *c* fuera encontrado experimentalmente en 1974, la teoría de Glashow, Weinberg y Salam fue aceptada ampliamente y sus autores reconocidos con el Premio Nobel en 1979, al igual que Carlo Rubbia y Simon van der Meer en 1984 por los experimentos en el CERN en 1983 que encontraron los bosones $Z^0$, W+ y $W_-$, y Veltman y 't Hooft en 1999 por probar la renormalización de la teoría.

## La interacción fuerte en el Modelo Estándar: la Cromodinámica Cuántica

Aquí retomamos la historia de los quarks y su nueva carga adicional con tres valores, propuesta en 1964 y 65. En 1971 Harald Fritzsch y Murray Gell-Mann llamaron a esta carga “color” como complemento del “sabor” de los quarks, y en 1972 Fritzsch, Gell-Mann y Heinrich Leutwyler, propusieron que, así como la interacción electrodébil también la fuerte fuera descrita como una teoría de Yang y Mills cuyos bosones mediadores, los “gluones” (del inglés “glue”, pegamento) no tienen masa y llevan cargas de color. La teoría relativista de campos cuánticos de quarks y gluones se llama Cromodinámica Cuántica (de la palabra griega “cromos”, color) o QCD (Quantum Chromo-Dynamics).

Sólo en 1973, David Politzer e independientemente David Gross y Frank Wilczek probaron que la QCD explica las propiedades de los quarks y los hadrones, y fueron reconocidos por ello con el Premio Nobel en 2004. Mostraron que, debido a que los gluones tienen carga de color, la interacción entre quarks aumenta de intensidad con la distancia, de modo que los quarks son casi libres a distancias muy cortas (propiedad que se llama “libertad asintótica”), pero no pueden separase a una distancia mayor del tamaño de un hadrón. Aunque los gluones tienen masa cero, como el fotón, están confinados dentro de objetos (los hadrones) que tienen carga de color cero: por lo tanto, no hay interacciones de color de larga distancia.

## El origen de la masa en el Modelo Estándar

El concepto de masa fue introducido por Newton en su famosa ecuación $F = ma$, para indicar la “inercia” de un cuerpo, que es su resistencia a cambiar su estado de movimiento. La masa $m$ aparece como un atributo intrínseco de un objeto que se define operacionalmente (usando la misma ley de Newton se mide la masa de un cuerpo relativa a la de otro definida como unidad). Esta es la definición de masa de objetos macroscópicos. En cambio, en el ME la masa de las partículas es una propiedad que depende de la interacción de estas con un campo bosónico, que “rompe espontáneamente” la simetría electrodébil.

Se dice que una simetría de un sistema esta “espontáneamente rota” cuando el sistema está en un estado en el que la simetría no es evidente. Un ejemplo típico es la magnetización de un imán de hierro. Cada átomo de hierro es como un pequeño imán. A altas temperaturas los átomos vibran mucho, por lo que los imanes atómicos apuntan en todas direcciones, la magnetización total es cero y la simetría esférica (bajo rotaciones en todas direcciones) del material es aparente. Al bajar la temperatura, los imanes atómicos de hierro empiezan a alinearse y finalmente por debajo de una temperatura critica, apuntan todos en una misma dirección elegida al azar, por lo que aparece una magnetización espontanea macroscópica del pedazo entero de hierro (ver la figura 13.a) y la simetría esférica dentro de este no es mas evidente.

Un ejemplo mucho más simple de rotura espontanea de simetría es el de una mesa redonda dispuesta para muchos comensales en la que cada vaso es equidistante de dos platos. Cuando un comensal elige el vaso de la izquierda (o el de la derecha), todos los otros por cortesía se ven forzados a hacer lo mismo. La simetría entre una “mesa izquierda” y una “mesa derecha” está rota espontáneamente una vez que un comensal elige un vaso.

La superconductividad es otro ejemplo más relevante. Si un material se encuentra en un campo magnético ($B$ en la figura 13.c) y se vuelve superconductor (al bajar la temperatura por debajo de un cierto valor critico), expulsa las líneas del campo casi totalmente. Este es el “efecto Meissner”. El campo magnético sólo penetra dentro de superconductor en una capa muy fina, cuyo espesor es el alcance del campo electromagnético dentro del material. Esto significa que el fotón adquiere una masa inversamente proporcional a ese alcance dentro del material. El fenómeno de la superconductividad, como fue explicado por Bardeen, Cooper y Schrieffer en 1957, se debe a la formación un condensado de “pares de Cooper” (cada par consiste de dos electrones ligados formando un bosón) y es dentro de este condensado que el fotón adquiere una masa (proporcional a la función de onda del condensado).

Yochiro Nambu en 1960 y luego Nambu y Giovanni Jona-Lasinio en 1962 propusieron un mecanismo de formación de pares similar al de la superconductividad para dar masa a los hadrones, pero no en una teoría de gauge. En 1962, por analogía con el efecto Meissner, Philip Anderson, un físico de materia condensada, mantuvo que también en teorías de campos la rotura espontánea de la simetría de gauge llevaría a tener bosones de gauge con masa. Pero el articulo de Anderson usaba un lenguaje no familiar para los físicos de partículas y sus ejemplos concretos eran no relativistas.

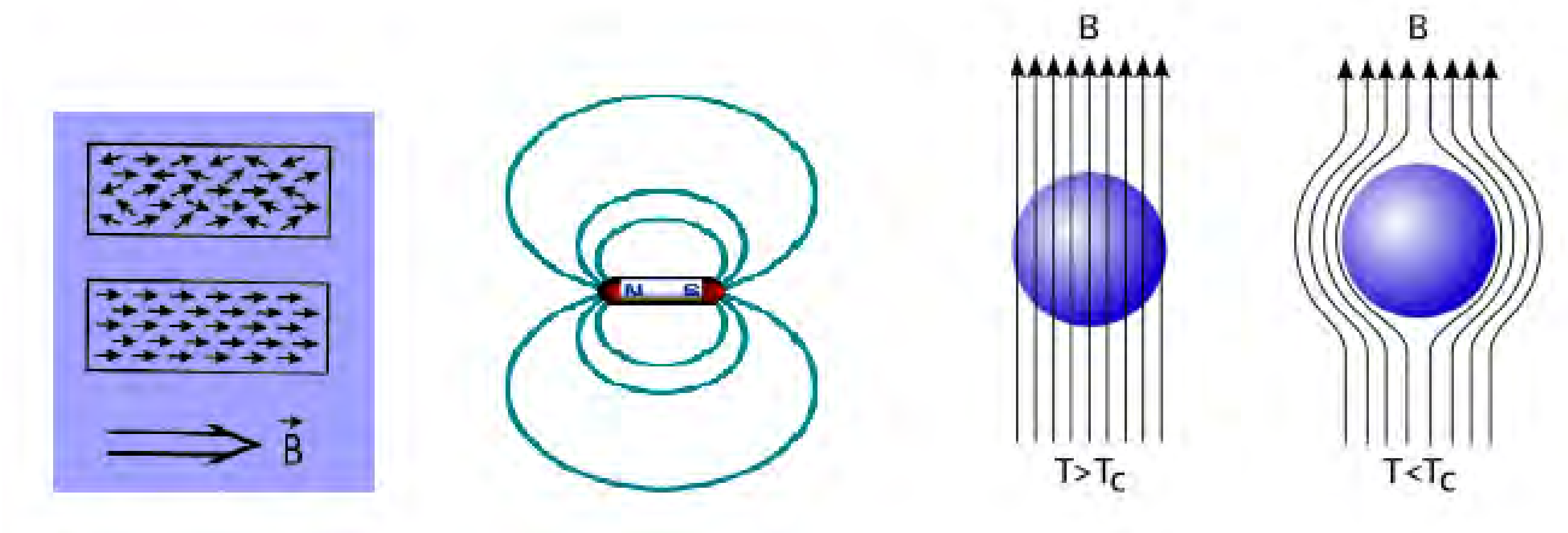

Figura 13: 13.a (izquierda) El panel superior muestra el material ferromagnético sin magnetización a altas temperaturas y el inferior, el material con todos los imanes atómicos orientados en la misma dirección, lo que produce una magnetización espontánea no nula, como la del imán mostrado en 13.b (centro izquierda) con las líneas de Faraday de su campo magnético. 13.c (derecha) Efecto Meissner: las líneas del campo magnético $B$ que pasan por dentro de un material (a temperaturas más altas que una temperatura crítica, $T \geq T_c$) son expulsadas cuando el material se vuelve superconductor (a temperaturas inferiores a la critica $T \leq T_c$ ).

La prueba de que el que ahora llamamos “mecanismo de Brout, Englert, Higgs (BEH)” o simplemente “mecanismo de Higgs” vale para teorías relativistas de campos cuánticos, fue dada en 1964 en tres artículos independientes y casi simultáneos: el primero de Francois Englert y Robert Brout, el segundo de Peter Higgs y último de Gerald Guralnik, Carl Richard Hagen y Tom Kibble [7]. Por los dos primeros artículos, Englert y Higgs han recibido el premio Nobel 2013 (que llegó demasiado tarde para Brout, fallecido en 2011) [8]. En estos artículos es por primera vez un campo bosónico elemental el que rompe espontáneamente la simetría. De los tres, sólo el artículo de Higgs menciona algunas propiedades de la partícula asociada con el campo bosónico, el ``bosón de Higgs”, y es por esto (y algo de suerte [7]) que es el nombre de Higgs el que se asocia con el mecanismo, el campo y el bosón. Ninguno de estos tres artículos aplicó el mecanismo al ME (sino Weinberg y Salam más tarde, como ya dijimos).

En el ME la rotura espontánea que da masa a las partículas elementales se debe al valor (técnicamente a un “valor de expectación en el vacío” o VEV) no nulo del campo de Higgs adicionado al modelo de Glashow ad-hoc por Weinberg y Salam. La aparición de un VEV no nulo es equivalente a la formación de un condensado de bosones en superconductividad. Las partículas adquieren una masa proporcional al VEV y a la magnitud de su interacción con el campo de Higgs. El fotón y los gluones no interactúan con este campo, que es eléctricamente neutro y no tiene carga de color, y por lo tanto no adquieren masa. Los tres bosones de la interacción débil, en cambio, así como los leptones y quarks sí interactúan con el campo de Higgs y, por lo tanto, adquieren masa. Una descripción intuitiva (ilustrada en la figura 14) es que en el VEV las partículas que interactúan con éste no pueden moverse a la velocidad de la luz, como partículas sin masa, sino a velocidades cada vez menores (lo que corresponde a que tengan una masa mayor) cuanto más se acoplen al campo de Higgs.

En 1993 el entonces Ministro de Ciencias británico ofreció una botella de champagne como premio a quien pudiera explicarle mejor el mecanismo de Higgs. Lo ganó David Miller de *University College*, *London*, quien ideó la analogía entre el condensado de Higgs y un salón lleno de personas, supongamos físicos (que representan los bosones en el condensado) que se muestra en la figura 15. Al llegar una persona famosa (como Einstein), los físicos se aglomeran a su alrededor tratando de verla y hablarle y esta se desplaza no como una persona sola sino como un gran grupo, que tiene mucha más dificultad para cambiar su velocidad, es decir, adquiere inercia, o sea, masa. En cambio, una persona desconocida podrá atravesar el salón sin mayor dificultad, ya que los físicos no se inmutarán.

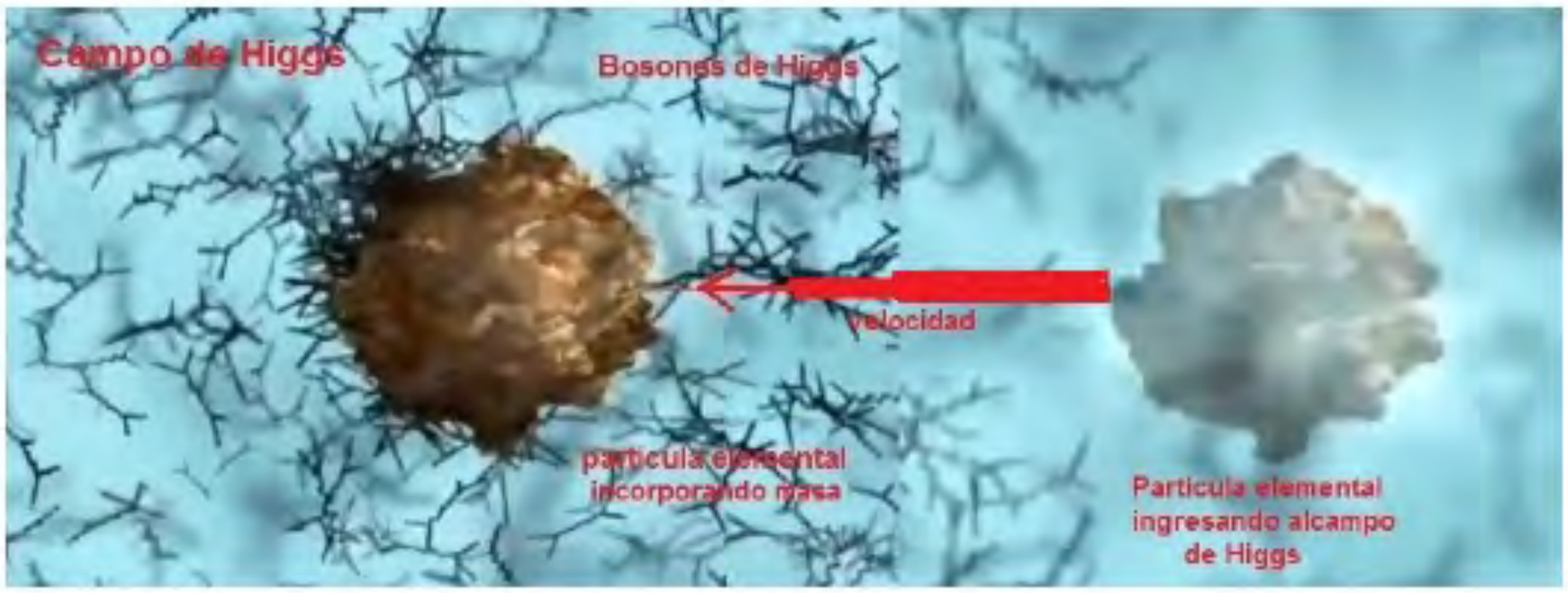

Figura 14: Interpretación artística de una partícula que adquiere masa en el “condensado” del campo de Higgs

Es importante mencionar que un bosón de gauge sin masa sólo tiene dos estados (o polarizaciones) posibles, pero uno masivo tiene tres. El estado adicional, manifiesta una componente adicional del bosón de gauge masivo que viene del campo bosónico que origina su masa. En el ME este campo tiene originalmente cuatro partes (o componentes), tres de ellas se asocian a (son “comidas” por) los tres bosones de gauge que adquieren masa y sólo una queda como un campo independiente, el campo de Higgs.

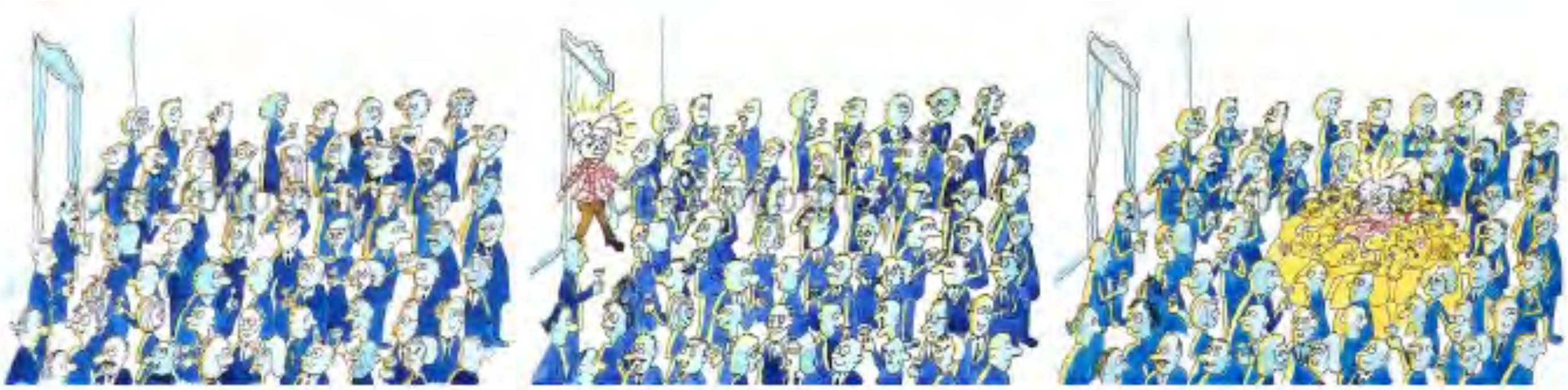

Figura 15: Explicación del mecanismo de Higgs para generar masa por analogía con una persona famosa que atraviesa un salón lleno de gente (dibujos obtenidos en el CERN Document Server).

La única manera de probar que es en verdad el mecanismo de Higgs el que opera en el ME era detectar el bosón de Higgs. De aquí la monumental importancia del descubrimiento de esta partícula. El problema fundamental para encontrarla fue que su masa no se puede predecir porque también se debe al VEV de su campo y a la intensidad de su auto-interacción, y ésta no está fija en la teoría. Una vez medida la masa del bosón de Higgs podemos predecir con exactitud dentro del

ME la intensidad de todas sus interacciones, lo que permitirá probar en el LHC en los próximos años si la partícula descubierta es en verdad este bosón.

Debemos aclarar en este punto, que la masa adquirida por el mecanismo de Higgs de los quarks *u* y *d* dentro de un neutrón o un protón sólo constituye aproximadamente dos centésimos de la masa de estos. El 98% restante de la masa de los átomos, y de toda la materia visible, se debe a las interacciones de la QCD [9]. Dentro de un hadrón, un quark interactúa con otros emitiendo gluones virtuales, que, como tienen color, pueden ellos mismos emitir otros gluones y pares de quark y antiquark, que generan otros gluones, etc. Además, hay pares de quark-antiquark que se ligan formando bosones (similares a los pares de Cooper) y forman un condensado dentro del hadrón. Esta nube de partículas virtuales y el condensado de pares quark-antiquark, resiste el movimiento acelerado de los quarks dentro de un hadrón y produce inercia, o sea, una masa que (para distinguirla de la masa del quark generada por el mecanismo de Higgs) se llama "masa constituyente". Para los quarks *u* y *d* esta masa constituyente es aproximadamente un tercio de la masa de un protón o neutrón. La masa constituyente es mucho menos importante para los quarks más pesados, porque su masa generada por el mecanismo de Higgs es similar o mucho mayor.

El cálculo de la masa del neutrón, del protón y de otros hadrones, empezando por primeros principios, basado sólo en la QCD, fue completado recién en el 2008 usando una red de supercomputadoras en Alemania, Hungría y Francia [8].

## Después del Modelo Estándar

El ME deja muchas cosas por explicar. Deja separadas la interacción fuerte y la electrodébil, no incluye la interacción gravitatoria, ni una explicación de la jerarquía de las masas de los leptones y quarks, sea su repetición en tres generaciones, sea la jerarquía dentro de cada generación. Tampoco da una explicación de la carga relativa del electrón y los quark, que asegura que los átomos sean neutros, y de por qué la materia atómica estable en el Universo está hecha sólo de átomos y no de anti-átomos.

Finalmente, el tipo de materia compuesta por átomos constituye sólo menos del 5% del contenido del Universo. Un 23% del contenido está en la forma de "materia oscura", un tipo de materia que no absorbe ni emite luz de ningún tipo y para la cual no hay candidatos en el ME, y el 72% restante está en una forma mas exótica todavía, la "energía oscura", que, contrariamente a la materia, tiene interacciones gravitacionales repulsivas y para la cual no hay ningún modelo teórico todavía.

Los físicos de partículas elementales esperamos que pronto el ME sea superado por otro que explique más sobre las partículas elementales y sus interacciones y que en los próximos años los experimentos en el LHC marquen la vía a seguir.

## Reconocimientos:

Agradezco muchísimo la ayuda de Roberto C. Mercader en la producción de este artículo, que no existiría sin su participación. Este trabajo fue apoyado en parte por el subsidio DE-SC0009937 del US Department of Energy.

## Referencias